# On some peculiarities of cosmic γ – radiation in the energy range ~ $10^{14}$ - $10^{15}$ eV

T.T. Barnaveli (*), T.T. Barnaveli Jr, N.A. Eristavi, I.V. Khaldeeva

Andronikashvili Institute of Physics, Tamarashvili 6, Tbilisi, 0177, Georgia

(*) E-mail: mantex@caucasus.net ;    barnaveli@hotmail.com

The distribution of EAS hadronic component energy fluxes through the ionization calorimeter in the primary energy range ~ $3 \cdot 10^{13} – 10^{16}$ eV is considered. The EAS, carrying the zero or the minimal flux of the hadron component energy are extracted. The conclusion supports the existence of isotropic γ – radiation in the primary energy region $E_0$ ~ $1 \cdot 10^{14} – 2 \cdot 10^{15}$ eV. This radiation has a spectrum of a form, close to the bell-like one, with the maximum at $E_0$ ~ $2.2 \cdot 10^{14}$ eV and with an additional local maximum at $E_0$ ~ $1.6 \cdot 10^{15}$ eV .

## 1  Introduction

This work is a continuation of the investigations [1,2], where the peculiarities of the  primary cosmic radiation nuclear composition behavior and of EAS hadron component energy fluxes in the energy range ~ $3 \cdot 10^{13} – 10^{16}$ eV were studied.

In this work the special accent is made on the extraction of EAS, carrying the zero or the minimal flux of the hadron component energy, i.e. being the candidates for γ – showers. The general character and the possible mechanisms of γ – radiation formation in the primary energy region ~ $10^{15}$ eV are analyzed for example in [3].

The data were obtained by means of Tian-Shan high mountain installation in the period before 1980 and were retreated anew in the frames of the new tasks and new approach. More than 400 000 events were handled in total. The data bank is described in [4], while the description of the installation can be found in [5], for example. The whole data treatment was carried out in the Andronikashvili Institute of Physics and in the company Mantex Computers Ltd. (Tbilisi).

## 2  The experiment

The main trigger of the installation ("shower trigger") was formed by scintillation detectors, registering the EAS electron – photon component. This trigger implied the imposing of certain requirements on the densities of particles in detectors, and on the relationship of these densities in the center and on the periphery of the installation. At the expense of this the lower boundary was set to the size $N_e$ of the registered EAS. This trigger caused the operation of the installation at $N_e$ >~ $2 \cdot 10^4$ particles.

It is essential, that at the same time the other triggers worked as well, without any special restrictions on the limits of the registered EAS diapason.

First of all, it was a so-called "calorimetrical" trigger, which caused the action of the installation at the energy deposit (the threshold was sufficiently small) in 3 upper layers of the multilayer ionization calorimeter (the total number of layers in this experiment was 16, the area of each – 36 $m^2$ ). This energy deposit could be caused both by hadron and by electron – photon components.

Besides that the so called "peripheral" trigger functioned, which caused the operation of the installation by action of the system of scintillation detectors, disposed in the form of concentric circles, with the center at the distance of about 50 meters from the center of installation.

These triggers did not impose any special conditions on the diapason borders of the registered EAS, though it is natural, that the registration efficiency decreases with the decrease of $N_e$. In this work we will consider only the EAS of the $N_e > 7 \cdot 10^3$ particles.

The events were selected with the distance R < 40 m of EAS axes from the center of the installation, for all of the triggers. The range of registration zenith angles was $0^o - 30^o$. The age parameter S was regarded as a free one, and could have the values in the limits 0.01 - 1.99, which does not contradict the mathematical structure of the Nishimura - Kamata function.

For each individual EAS $N_e$ and $E_h(N_e)$ were calculated, and then on the plane with coordinates $N_e$, $EK = E_h(N_e)/n$ for each event the corresponding point was plotted. Here n – is the number of discharged chambers in calorimeter, n >=1, while under $E_h$ the whole disposure of hadron component energy in the calorimeter is meant (starting from the 4-th layer and below). The result is shown in Fig.1.

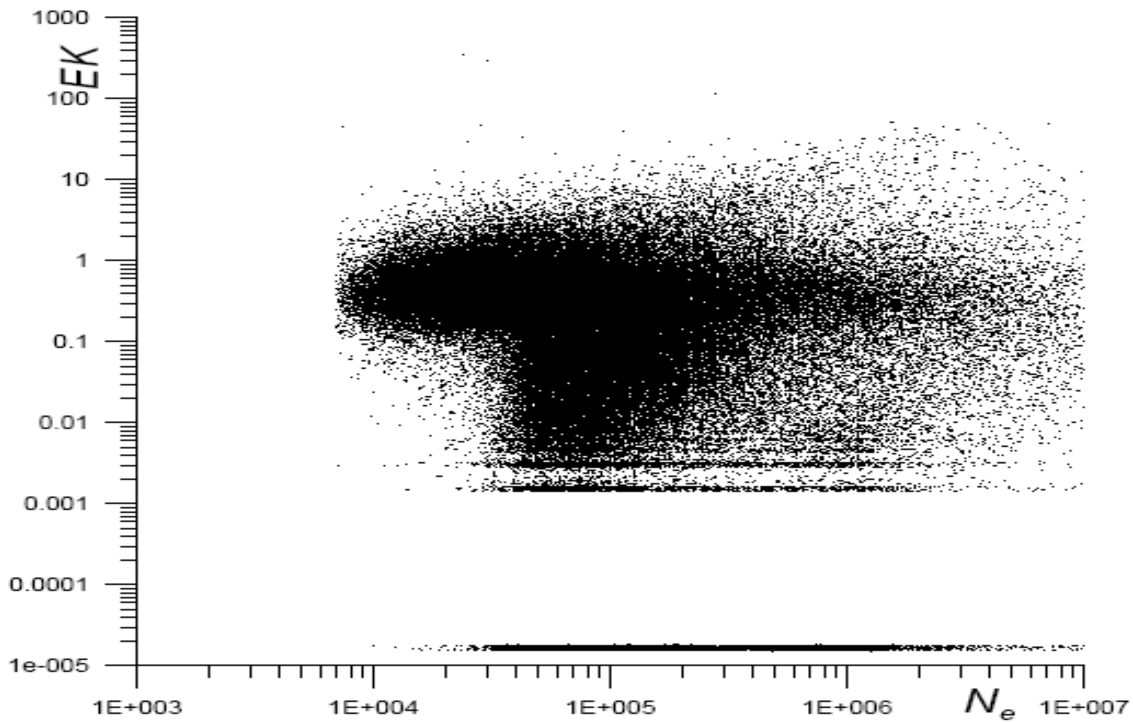

Fig. 1. The specific (per one chamber) disposure of hadron component energy in the calorimeter as function of $N_e$

Two basic structures can be resolved on this distribution with a sufficient observability. These structures were defined by the above indicated main triggers of the installation – the "calorimetrical" and the "shower" trigger. One of these structures is a broad strip "A" in the upper part of the graph, stretching parallel to the $N_e$ axis. In vertical its main part occupies the region $10^{-1} < EK < 10$. Due to the high steepness of the primary cosmic radiation (PCR) spectrum, the right hand part of this strip is represented rather diffusely, however it is possible to follow it. This structure is defined by the events, selected by the "calorimetrical" trigger, which required the deposition of the over-threshold energy at least in one of the very upper layers of the calorimeter. It is natural, that on the right side and on the top this aggregate of the points is not limited. The probability of the event registration on the left side depends on the probability of the chamber discharge in the upper layers of the calorimeter at small values of $N_e$. The second structure is the condensation "B" of points for the $N_e$ around ~ $10^5$. On its left side this structure is limited by the conditions of the "shower" trigger. On the right side and on the top it is not limited. By its shape and position it sharply differs from the structure "A". The main part of the events here is placed in the limits $3 \cdot 10^4 < N_e < 5 \cdot 10^5$ ($1 \cdot 10^{14}$ eV $< E_0 < 2 \cdot 10^{15}$ eV). As to "peripheral" trigger, the events defined by this trigger are distributed more evenly on the graph plane, and are not distinguishable against the background of the two first structures.

In the lower part of the graph, one can see several narrow strips, stretching parallel to $N_e$ axis. These strips (with the exception of the lowest one) correspond to the extraction of the minimal over-threshold energy in one, two, three and so on chambers of calorimeter in the individual events. The parallelism of the strips to $N_e$ axis is the result of standard energy (the minimal over-threshold) and standard number of the discharged chambers in each individual EAS. The width of the strips is determined by the interval of the registered zenith angles of events, since in each event the correction is introduced, proportional to cosine of the registration angle.

We tried to extract out of the whole aggregate of events, those of EAS which most probably are generated by primary γ - quanta. The only reliable signature of EAS γ – nature available at this time is the complete absence of hadrons in the calorimeter (in any case, of more or less energetic ones: as we said above, the energy release only below the 3-d layer of calorimeter is considered).

In the standard way of data handling, the release of the minimal over-threshold energy in at least 1 out of 624 chambers of 13 lower layers of calorimeter was required. Now, to provide the extraction of hadron-less showers, the extremely small, however nonzero energy $E_{conv}$ (by 4 orders smaller than the threshold energy of chambers) was ascribed to each chamber of calorimeter without exception. Thus the events containing no hadrons at all in the calorimeter turned out to be marked with the standard energy $E_{conv} \cdot n$, where n – is the number of chambers in the calorimeter.

The hadron-less events form the separate strip in Fig. 1, lying much lower than the rest of the events. The total number of events on the graph, in such way of data representation is ~ 122 000. The number of events in hadron-less strip is 5670. Note that these strips (and not only the lowest one) can emerge with the sufficient probability only in those regions of $N_e$, where there is enough amount of γ – EAS. All three triggers contribute to these strips. A little later, the contributions of different triggers will be considered separately. Of course, some part of γ – EAS still contains a large amount of nuclear active component, imitating the ordinary EAS, generated by nuclei. The hadron-less strip will be shown later on a large scale.

Let us consider some of the properties of the structures "A" and "B".

The first, that attracts the attention - is a strict distinction of these structures in the flux of hadron energy through the calorimeter. These fluxes in the structure "B" are essentially lower (down to 2 orders, on average by 1 order) compared with corresponding fluxes in structure "A". It is natural, because the events of the structure "A" are defined by the "calorimetrical" trigger and contain the hadrons with much higher probability. Due to the extremely small flux of nuclear active component in the events forming the lower part of structure "B", the natural supposition will be, that the essential part of these events is generated by γ – quanta of the energy around ~ $2.2 \cdot 10^{14}$ eV. Note, that except for the small flux of the hadron component in these events, the behavior of age parameter S speaks in favor of γ – nature of the part of these showers. For the EAS constituting the structure "B", the age parameter S is essentially higher than for EAS from the basic structure "A". This allows one to conclude, that the translation of the energy by the leading particle, which is characteristic for nuclear cascade, is not characteristic for these events, i.e. the energy of these EAS dissipates faster than in EAS initiated by primary hadrons and nuclei. So, at the level of observation, these showers have the relatively higher value of age parameter S.

As we said above, in this article the special attention is paid to the hadron-less EAS.

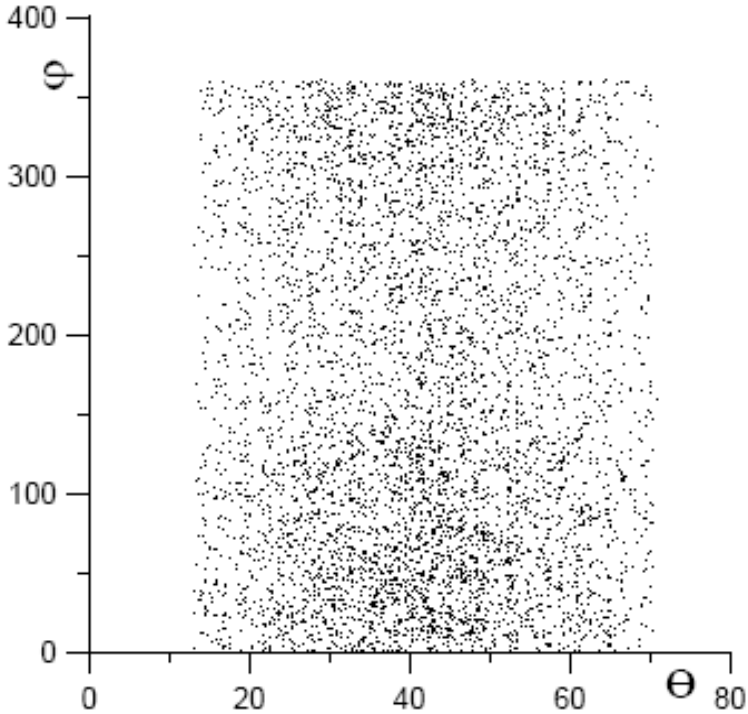

Fig 2. The distribution of the astronomic coordinates
of the axes directions for the hadron-less EAS

Naturally the question arises, could not the events constituting the strip of hadron-less EAS, be the result of emission of some discrete source? In this case, one can expect the appearance of segregated parts on the distribution of the fall directions of EAS, which form this strip. The distribution of the astronomic coordinates of the EAS axes directions for this strip is given in Fig. 2.

The installation is located at ~ 43º north latitude. It is easy to see in the graph. No noticeable signs of any segregated coordinates can be seen. This precludes the supposition, that the strip of hadron-less EAS is the result of the radiation from some discrete source.

Let us consider now the distribution of events density in the lowest strip (of the completely hadron-less EAS) in Fig. 1, as a function of $N_e$. For the clearness of the picture, we divided all events in accordance with the trigger, that selected them. (It is natural, that some of the events were selected simultaneously by more than one trigger). These distributions are shown in Fig. 3: a) - "shower trigger", b) - "calorimetrical trigger" and c) - "peripheral trigger". The maximum of all shown distributions is at $Ne \sim (7.5 \pm 1) \cdot 10^4$. On the right wing of the distributions in Figs. 3 a) and b) in the region of $N_e \sim (8.5 \pm 1) \cdot 10^5$ one can easily see the presence of the additional irregularity.

We would like to underline especially, that to the left from the maximum, all these distributions behave in a similar way, despite the above mentioned presence or absence of limitations imposed on event registration by different triggering conditions.

The graph in Fig. 3 c) requires additional elucidation. The events constituting this graph were obtained at the "peripheral" trigger, demanding the operation of the detectors placed at distances up to 90 m. So the large amount of EAS, with the axes located at large distances from the calorimeter (close to 40 m – the upper limit of the permitted axes locations) was involved in the consideration. Thus, the contributions increase of the events, which are not the γ – EAS, but nevertheless are not accompanied by the registration of hadrons in the calorimeter. Due to this, the peculiarities of distribution are not revealed distinctly. Seemingly, the existence of the above considered peculiarities creates plateau-like upper part of the distribution. It is sited between the values of $N_e$, corresponding to these peculiarities.

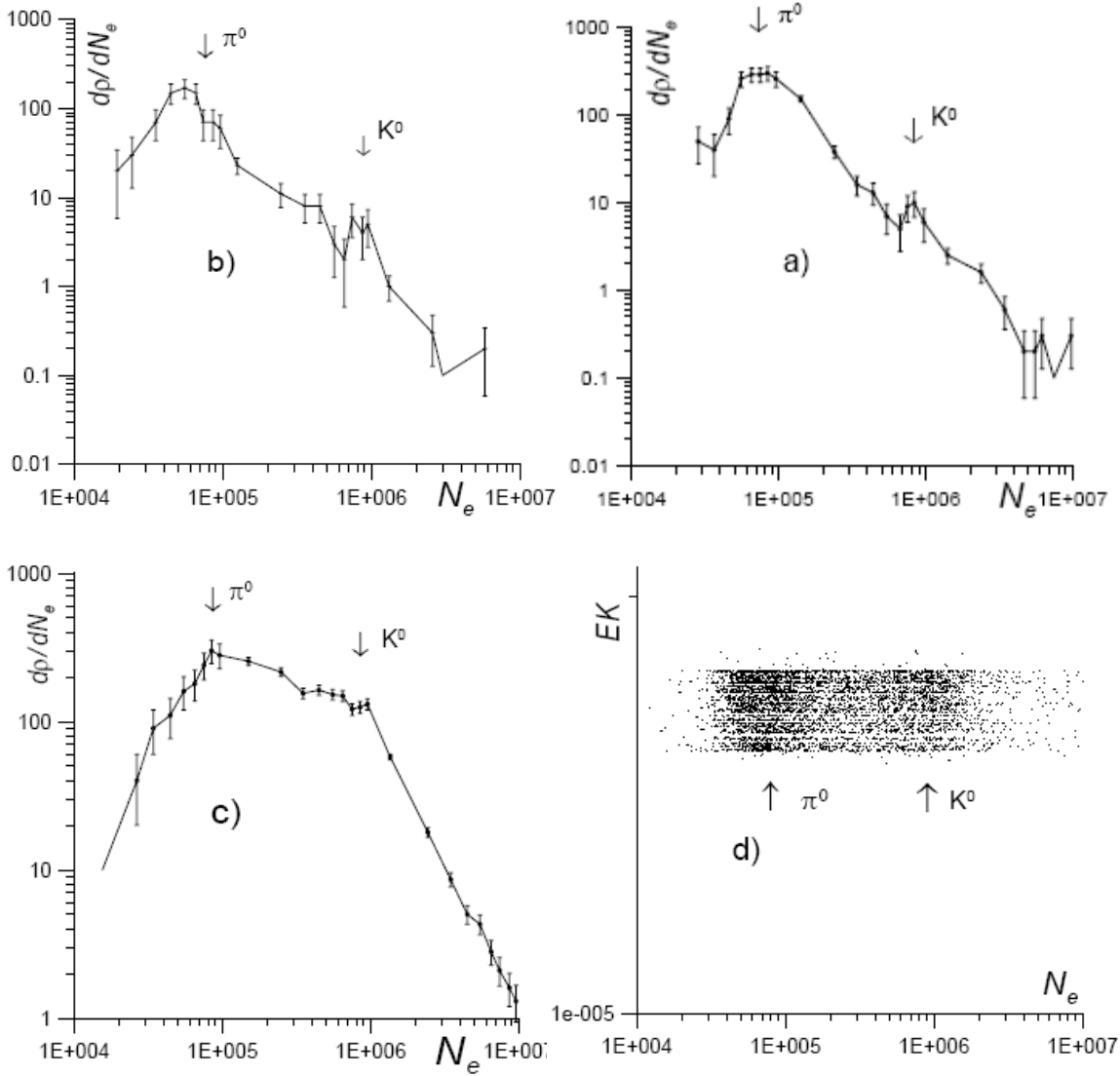

Fig. 3. The distribution of events density in the lowest strip (of the completely hadron-less EAS), as a function of $N_e$.

Again, the hypothesis was checked on the possible existence of discrete sources, which could be responsible for the generation of EAS, lying in the regions of the mentioned maximums of Figures 3 a), b) and c) (but not for the whole strip of hadron-less showers, as was done above). However the events, constituting each of these structures, again did not reveal any noticeable signs of anisotropy in directions of their axes.

If one accepts, that above considered hadron-less EAS are created by high energy γ – radiation, one can conclude, that this radiation is distributed in the region of primary energies $10^{14} - 3 \cdot 10^{15}$ eV, according to the law close to the bell - like one, with the maximum in the region of primary energies $E_0 \sim 2.2 \cdot 10^{14}$ eV ( $N_e \sim 7.5 \cdot 10^4$ ) and with an additional local maximum in the region $E_0 \sim 1.6 \cdot 10^{15}$ eV ( $N_e \sim 8.5 \cdot 10^5$ ).

The increasing of the image scale of the strips in the lower part of Fig. 1 leads to an interesting picture. The inner structure of these strips is revealed, connected in particular with the angular distribution of the registered events. In Fig. 3 d), the hadron-less EAS strip, created at joint action of all three above mentioned triggers is represented in increased scale (i.e. the lowest strip in Fig. 1). The zenith angles of the registered event axes increase upwards in this graph. The energy, released in the calorimeter in each event, is corrected according to the cosine of the registration angle. In the case of the above described standard signal marking the hadron-less EAS, this leads to the broadening of the line on the graph and its conversion into the strip. The uneven populating of the strip by vertical, reflects, by essence, the joint influence of the events angular distribution and of the acceptance of the installation under the different zenith angles. The value of this graph is, that here once again the above-mentioned peculiarities of the hadron-less EAS distribution are revealed clearly. They can be seen here as clearly visible vertical strips of point condensation around the $N_e \sim 7 \cdot 10^4$ and $N_e \sim 8 \cdot 10^5$, regardless of the event registration angles. Note that similar condensations of events can be seen on the strips next to the hadron-less one on the Fig. 1, and even in the area above them in the lower part of the graph. So these peculiarities take place for the spectra of EAS, carrying the minimal fluxes, and not only for zero fluxes of the hadron component energy.

The presence of this γ – radiation, its observed peculiarities and their localization on the primary energy scale quite naturally fits the frames of the model proposed in the works [1,2]. According to this model, interstellar space contains the isotropic cold background of light particles of the mass ~ 36.7 eV. ( It is not excluded, that this particle is one of the heavy neutrinos, most probably $\nu_\tau$, though this may contradict the recently accepted view concerning the expected neutrino masses). The interaction with this background leads to the cutoff of the primary nuclei spectra, due to their destruction [2]. The process has a threshold character, and is switched on starting from the energies depending on the masses of the nuclei, and on the nucleons bonding energy in them. For example, for He4 the threshold energy of destruction on this background is $2.02 \cdot 10^{15}$ eV, for Fe56 it equals $1.42 \cdot 10^{16}$ eV. However, the protons (like the nuclei) must interact with this background as well. As a result of this interaction, starting from the corresponding energies, $e^+e^-$, $\pi^0$, $\pi^+\pi^-$, $K^0$, $K^{+-}$, η-mesons etc. must be created.

Generation of $\pi^0$ – mesons at the interaction of primary protons with the isotropic cold background of light particles of the mass of ~ 36.7 eV, leads to the creation of γ – quanta of the energy ~ $2.17 \cdot 10^{14}$ eV in the laboratory system. The corresponding $N_e$ is ~ $7.7 \cdot 10^4$. This value is in good accordance with the location of the maximum in the Figs. 3 a), b) and c), and of the first condensation strip in Fig. 3 d).

The following is to be especially taken into account. If one accepts that for any nuclide, each of the nucleons interacts with the background particle independently from others, then the energies of the created γ – quanta for every nuclide will overlap naturally. Actually, since for the creation of $\pi^0$ the same γ – factor is required both for the initial nucleon or initial nuclide, so all the $\pi^0$ - s will have the same energy in the system of the initial nuclide, regardless of its mass. If one admits the possibility of background particle interaction with the α – particle (existing in the structure of the initial nuclide), so in this case the energy of $\pi^0$ ( and consequently the energy of generated γ – quanta) will increase by ~ 10% . Namely, $E_{\pi 0}$ becomes equal to $4.8 \cdot 10^4$ eV while the mean energy per one created γ – quantum will be ~ $2.4 \cdot 10^{14}$ eV. The corresponding $N_e$ is ~ $8.6 \cdot 10^4$.

As said above, in the same Fig.-s 3 a) and b) in the region $N_e \sim 8 \cdot 10^5$, one can observe one more peculiarity with the excess of the event density. It is easy to show, that in the frames of the considered model, this peculiarity by its origin is analogous to the previous one - and can be the result of $K^0$ creation in the interaction of the primary nucleon (or nucleon inside the nucleus) with the background particle. Here each of the created $\gamma$ – quanta (from 4 $\pi^0$ – s, since here are two $K^0$ due to strangeness conservation) carries the energy ~ $1.63 \cdot 10^{15}$ eV. The corresponding $N_e$ is ~ $8.5 \cdot 10^5$. This value as well is in a good accordance with the location of the second maximum in the Fig. 3 a) and b) and of the second condensation strip in Fig. 3 d).

Note, that the peaks of $\pi^0$ and $K^0$ must differ by several hundred times by their highs due to the steepness of the primary spectrum – they are shifted along the energy scale by about one order (due to the relationship of threshold energies of $\pi^0$ and $K^0$ generation).

## 3 Additional remarks

1. By the creation of η-mesons on the particles of the interstellar background (η-meson also has two-particle decay mode (~40%) with the generation of 2 $\gamma$ – quanta) each of the created $\gamma$ – quanta carries energy of ~ $2.6 \cdot 10^5$ eV. The corresponding number of particles in EAS $N_e$ is ~ $1.5 \cdot 10^6$. One can hope to observe the corresponding signal only by increasing of the statistics.

2. In general, in the graphs similar to the given above, any of the structures created by $\gamma$ – quanta generated in the threshold processes, must have steep left hand (low energetic) fronts. However, due to the existing probability of the detecting $\gamma$ – quanta from the very far galaxies, the left fronts of the mentioned structures must acquire the diffuse form due to red shift. Nevertheless, despite the sufficiently high precision of data handling, the statistic available to day is not enough for even the qualitative evaluation of such diffuseness, even if it really exists.

3. Attempting to extract the $\gamma$ – EAS one must take into account, that the hadron-less events may be imitated by events really containing hadrons - but having their axes at large distances from the calorimeter, since the hadrons are attracted mainly to the center of EAS. That is why the distances between the calorimeter and EAS axes, were from the very beginning restricted by the value < 40 m.

4. The behavior of the π-, K- and η – mesons generating cross sections (in the sense of the shift against the threshold) most probably is negligible, because one can accept that the cross section increases started from the threshold up to the maximum on the diapason of the order of the generated particle mass.

## 4 Conclusions

1. The distribution of hadron component energy fluxes in the individual EAS in the particle number diapason $N_e > 7 \cdot 10^3$ is constructed. The total number of the handled events is ~ 400000; the number of EAS contributing to the final result is ~ 122000.

2. On the basis of this distribution, EAS were selected which do not contain the hadron component at all, and which, with high probability, are representing the $\gamma$ – EAS. The number of such events is 5670.

3. The spectrum of these $\gamma$ – EAS in the region $3 \cdot 10^4 < N_e < 5 \cdot 10^6$ is constructed. The corresponding primary energies are $1 \cdot 10^{14}$ eV $< E_0 < 7.4 \cdot 10^{15}$ eV. The spectrum has a form close to the bell-like one with the maximum at $E_0$ ~$2.2 \cdot 10^{14}$ eV ( $N_e$ ~ $7.5 \cdot 10^4$ ) and with the additional maximum at $E_0$ ~$1.6 \cdot 10^{15}$ eV ( $N_e$ ~ $8.5 \cdot 10^5$ ).

4. The presence of this $\gamma$ – radiation, its observed peculiarities and their localization on the primary energy scale, quite naturally fits the frames of the model proposed in the works [1,2].

### Acknowledgments

The authors express their deep gratitude to O.V.Kancheli for his constant attention to our investigations and for fruitful discussions and remarks. The authors are sincerely grateful to V.P.Pavluchenko and V.I.Iakovlev for useful discussions concerning the data bank and installation, A.D.Erlikin and A.Wolfendale - for their interest in these investigations, J.M.Henderson for his interest in our investigations and his assistance in preparation of this paper.

This investigation is in part supported by a grant from the Georgian Academy of Sciences and by the firm Manex Computers ltd. (Tbilisi).